\documentclass{PoS}

\usepackage{amsmath}
\usepackage{here}
\usepackage{graphicx}
\usepackage{epsfig}
\usepackage{subcaption}

\def\nin{\noindent}
\def\beq{\begin{equation}}
\def\eeq{\end{equation}}
\def\bea{\begin{eqnarray}}
\def\eea{\end{eqnarray}}
\def\nnb{\nonumber}

\def\Eq#1{Eq.~(\ref{#1})}
\def\td{{\widetilde \delta}}

\title{First Numerical Implementation of the Loop--Tree Duality Method}

\ShortTitle{Numerical Implementation of the LTD}

\author{\speaker{Sebastian Buchta}\\
        Instituto de F\'{\i}sica Corpuscular, Universitat de Val\`{e}ncia 
- Consejo Superior de Investigaciones
Cient\'{\i}ficas, Parc Cient\'ific, E-46980 Paterna, Valencia, Spain \\
        E-mail: \email{sebastian.buchta@ific.uv.es}}
        
\abstract{The Loop-Tree Duality (LTD) is a novel perturbative method in QFT that establishes a relation between loop--level and tree--level amplitudes, which gives rise to the idea of treating them simultaneously in a common Monte Carlo. Initially introduced for one--loop scalar integrals, the applicability of the LTD has been expanded to higher order loops and Feynman graphs beyond simple poles. For the first time, a numerical implementation relying on the LTD was realized in the form of a computer program that calculates one--loop scattering amplitudes. We present details on the employed contour deformation as well as results for scalar and tensor integrals.}

\FullConference{The European Physical Society Conference on High Energy Physics\\
		22--29 July 2015\\
		Vienna, Austria}

\begin{document}

\section{Introduction}
\nin
When calculating NLO (NNLO) cross-sections one needs to consider the tree- and loop-contributions separately. Especially loops with many external legs prove to be challenging. Considerable progress has already been made in order to attack this problem: OPP- Method, Unitarity Methods, Mellin-Barnes Representation, Sector Decomposition \cite{Gehrmann:2010rj}. Still a lot of effort has to be put in to cancel infrared singularities among real and virtual corrections. Additional difficulties arise from threshold singularities that lead to numerical instabilities. 
The LTD method aims towards a combined treatment of tree- and loop- contributions. Therefore the LTD method casts the virtual corrections in a form that closely resembles the real ones \cite{Catani:2008xa,Bierenbaum:2010cy,Bierenbaum:2012th,Buchta:2014dfa,Hernandez-Pinto:2015ysa,Buchta:2015xda,Buchta:2015wna}.
\section{Loop-Tree Duality at one loop}
\begin{figure}[h]
  \centering
  \includegraphics[scale=1]{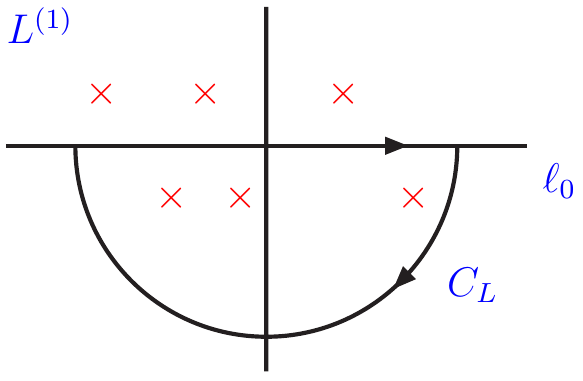}
  \caption{Location of poles and integration contour $C_L$ in the complex $\ell_{0}$-plane.}
    \label{fig:contour}
  \end{figure}

\nin
The most general, dimensionally regularized one-loop scalar integral can be written as \cite{Catani:2008xa}
\beq
L^{(1)}(p_1, p_2,\dots , p_N) = \int\limits_{\ell}\prod\limits_{i=1}^NG_F(q_i)~,
\label{eq:olamp}
 \eeq
  with the Feynman propagator $G_F(q_i) = [q_i^2-m_i^2+i0]^{-1}$, internal momenta $q_i = \ell + p_1 + \dots + p_i = \ell + k_i$ and shorthand integral notation $\int_{\ell} = -i\int d^d\ell/(2 \pi)^d$. As a first step, one performs the integration over the complex energy components of the loop four-momentum by applying the residue theorem. The integration contour is chosen such that it encloses the poles with positive energy and negative imaginary part, see Figure \ref{fig:contour}.
The residue theorem is employed by taking the residues of the poles inside of the contour and summing over them. Given an appropriate gauge choice the integrand in \Eq{eq:olamp} contains only simple poles. This leads to the dual representation of the one-loop scalar integral
\beq
L^{(1)}(p_1,p_2,\dots ,p_N)=-\sum\int_{\ell}\tilde{\delta}(q_i)\prod\limits_{\substack{j=1\\ j\neq i}}^NG_D(q_i;q_j)~,
\eeq
with $\tilde{\delta}(q_i)=2\pi i\delta_+(q_i^2-m_i^2)$ and $G_D(q_i;q_j)=1/(q_j^2-m_j^2-i0\eta(q_j-q_i))$.
The subscript ``+'' of the delta function indicates that the positive-energy solution is to be taken. Furthermore, $\eta$ is a future-like vector, i.e. $\eta^2\geq 0$, $\eta_0 > 0$. It is dependent on the choice of coordinate system, however it cancels out once one adds all dual contributions. Hence physical objects like scattering cross sections will stay frame-independent. Evaluating the ``non-cut'' propagators at the position of the pole leads to the so called ``dual prescription'' which serves to keep track of the correct sign of the i0-prescription of the corresponding propagator. Thus, by virtue of employing the residue theorem, it is possible to rewrite a one-loop amplitude as a sum of single-cut phase-space integrals over the loop-three-momentum. The i-th dual contribution has the i-th propagator set on-shell while the left over Feynman propagators get promoted to dual propagators.
  \begin{figure}[H]
  \centering
  \begin{subfigure}[h]{0.3\textwidth}
  \beq
L^{(1)}(p_1,p_2,\dots ,p_N)=\nnb
\eeq
\end{subfigure}
\begin{subfigure}[h]{0.69\textwidth}
  \includegraphics[scale=1]{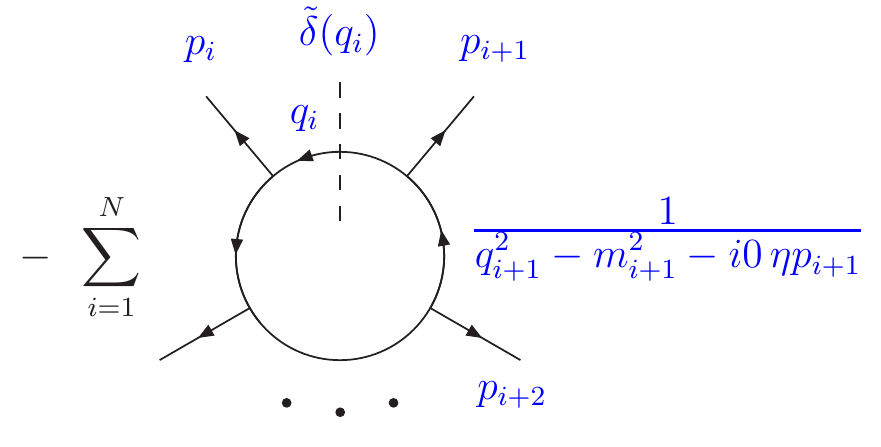}

    \end{subfigure}
      \caption{Graphical representation of the solution of the LTD at one-loop.}
  \end{figure}
The LTD features a set of interesting properties: The number of single cut dual contributions equals the number of legs, this way a loop diagram is fully opened to tree diagrams. Singularities of the loop diagram appear as singularities of the dual integrals. Tensor loop integrals and physical scattering amplitudes are treated in the same way since the LTD works only on propagators. Virtual corrections are recast in a form, that closely parallels the contribution of real corrections.
This is the formalism for the one-loop case. Solutions for more complicated situations like multiple loops \cite{Bierenbaum:2010cy} or higher order poles \cite{Bierenbaum:2012th} are described in the respective references.

\nin
\section{Singular behavior of the loop integrand}
\begin{figure}[h]
  \centering
  \includegraphics[scale=0.75]{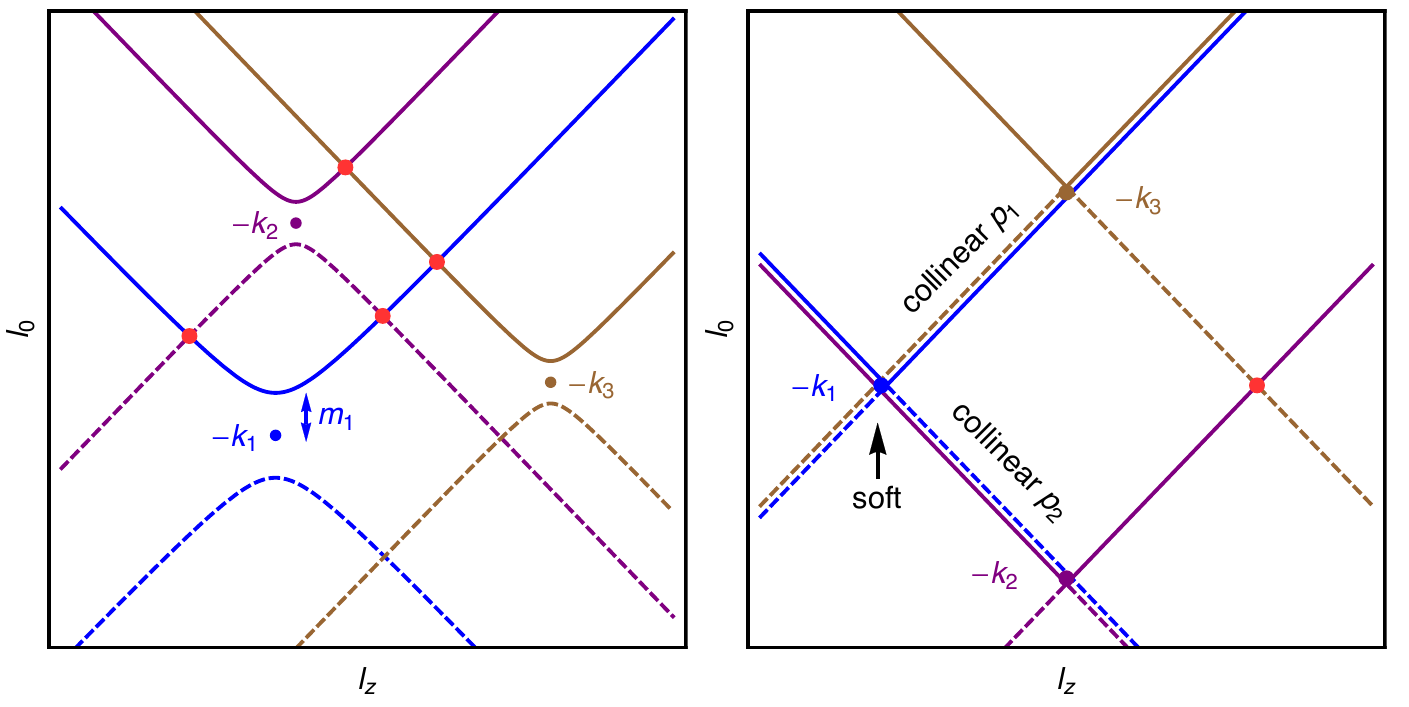}
  \caption{On-shell hyperboloids for three arbitrary propagators in Cartesian coordinates. The massive case is shown on the left and the massless on the right.}
    \label{fig:lightcones}
  \end{figure}

\nin
As a preparatory step it will prove useful to introduce an alternative way of denoting the dual propagator. This will give a more natural access to its singularities \cite{Buchta:2014dfa}.
\beq
\tilde{\delta}(q_i)G_D(q_i;q_j)=2\pi i\frac{\delta(q_{i,0}-q_{i,0}^{(+)})}{2q_{i,0}^{(+)}}\frac{1}{(q_{i,0}^{(+)}+k_{ji,0})^2-(q_{j,0}^{(+)})^2}~,
\label{eq:rwprop}
\eeq
with $k_{ji}=q_j-q_i$ and $q_{i,0}^{(+)}=\sqrt{\mathbf{q_i}^2+m_i^2-i0}$. 
In Fig. \ref{fig:lightcones}, the on-shell hyperboloids of three propagators in loop-mometum-space are sketched.
The loop integrand becomes singular at hyperboloids with $q_{i,0}^{(+)}=\sqrt{\mathbf{q_i}^2+m_i^2-i0}$ (solid lines) and $q_{i,0}^{(-)}=-\sqrt{\mathbf{q_i}^2+m_i^2-i0}$ (dashed lines) and origin in $-k_{i,\mu}$. Applying the LTD is equivalent to integrating along the \textit{forward on-shell hyperboloids}. The intersection of two forward hyperboloids (solid lines) leads to singularities that will cancel among dual integrals. The intersection of a forward with a backward hyperboloid (solid line with dashed line) leads to a singularity that remains and therefore has to be dealt with by contour deformation.\\
Singularities appear where the denominator of the dual propagator vanishes. Since the denominator in \Eq{eq:rwprop} has been rewritten as a difference of squares, we can easily extract two conditions for which the singularities show up:
\bea
q_{i,0}^{(+)}+q_{j,0}^{(+)}+k_{ji,0}=0\label{eq:condel}~,\\
q_{i,0}^{(+)}-q_{j,0}^{(+)}+k_{ji,0}=0\label{eq:condhyp}~.
\eea
\Eq{eq:condel} describes an ellipsoid in the loop three-momentum and demands $k_{ji,0}<0$. An ellipsoid is the result of the intersection of a forward with a backward hyperboloid. The origins of the hyperboloids are separated in a time-like fashion, expressed by the conditions
\beq
k_{ji}^2-(m_j+m_i)^2\geq 0, \quad k_{ji,0}<0~.
\eeq
The singularity described by eq. \Eq{eq:condhyp} has a hyperboloid shape as a result of the intersection of two forward on-shell hyperboloids of space-like separation. The corresponding condition is
\beq
k_{ji,0}^2-(m_j-m_i)^2\leq 0~~.
\eeq
Here, $k_{ji,0}$ may be positive or negative.
\nin
\section{Numerical Implementation}
\nin
The singularities associated with ellipsoid intersections require contour deformation. The following one-dimensional example illustrates how this can be achieved. The function
\beq
f(\ell_x)=\frac{1}{\ell_x^2-E^2+i0}~,
\eeq
has poles at $\pm (E-i0)$. Simply integrating along the real axis leads to infinities. Therefore the integration contour has to be deformed to go around the poles. A suitable contour deformation is
\beq
\ell_x\rightarrow\ell_x'=\ell_x+i\lambda\ell_x\exp\left(-\frac{\ell_x^2-E^2}{2E^2}\right)~.
\label{eq:cdef2d}
\eeq
The parameter $\lambda$ serves to scale the deformation along the imaginary axis. At the position of the pole, the exponential function reaches its maximum. Far away from the poles, the exponent is a large negative number, hence exponentiating it suppresses the deformation.
For the integration in loop three-momentum-space \Eq{eq:cdef2d} needs to be generalized to three dimensions. This is done by modifying the exponent and, of course, promoting $\ell_x$ to $\vec{\ell}$
  \beq
  \vec{\ell}\rightarrow\vec{\ell}'=\vec{\ell}+i\lambda\vec{\ell}\exp\left(-\frac{G_D^{-2}}{\text{A}}\right)~.
  \eeq
The parameter $A$ serves to control the width of the deformation. 
    \begin{figure}[H]
  \centering
\includegraphics[scale=0.8]{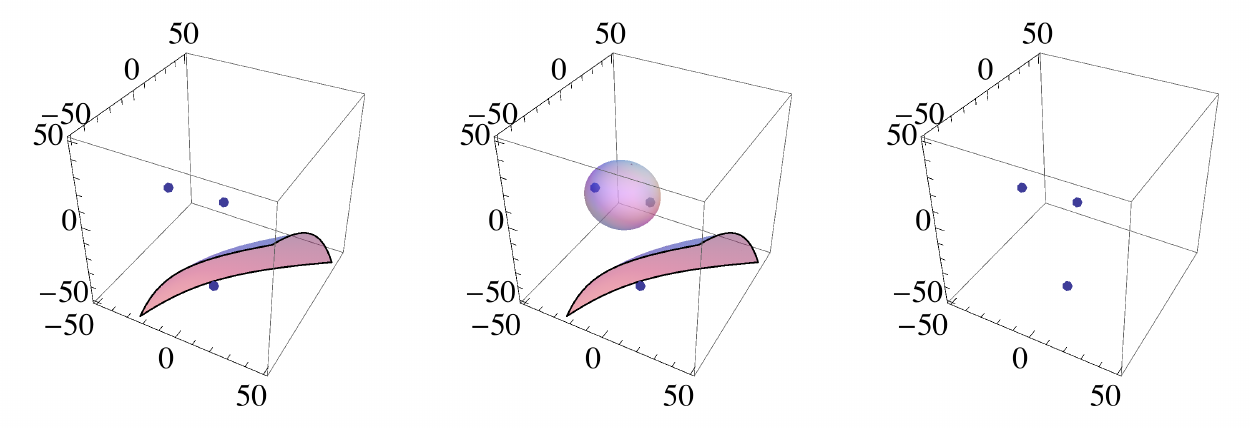}
  \caption{Hyperboloid and ellipsoid singularities in a triangle.}
    \label{fig:dualcont}
  \end{figure}
  The example in Fig. \ref{fig:dualcont} shows the location of the singularities as they appear in a triangle. Every box corresponds to a dual contribution. The easiest case is the third box, which doesn't display any singularities of any type and thus can be integrated straightforwardly. In the first contribution there is only a hyperboloid singularity, which cancels with the hyperboloid singularity from the second contribution. But since the second contribution features an ellipsoid singularity as well, this contribution requires contour deformation. In turn, in order to \textit{preserve the cancellation} of the hyperboloid singularities, contribution one will need the exact same deformation, \textit{despite} no ellipsoid being present there. This means that hyperboloid singularities act as a coupling between different dual contributions. The coupled contributions need to be deformed according to all ellipsoid singularities which they \textit{both} have.\\
  In the massless case collinear singularities arising from forward-forward intersections cancel among dual integrals, because the on-shell hyperboloids of Fig. \ref{fig:lightcones} are tangential there. Nonetheless, collinear and soft singularities from forward-backward intersections remain, but they are restricted to a finite region that can be mapped to the real phase-space emission \cite{Buchta:2014dfa,Hernandez-Pinto:2015ysa}.
  \nin
\section{Results and Conclusion}
\begin{figure}[h]
  \centering
  \begin{subfigure}[h]{0.49\textwidth}
\includegraphics[scale=0.62]{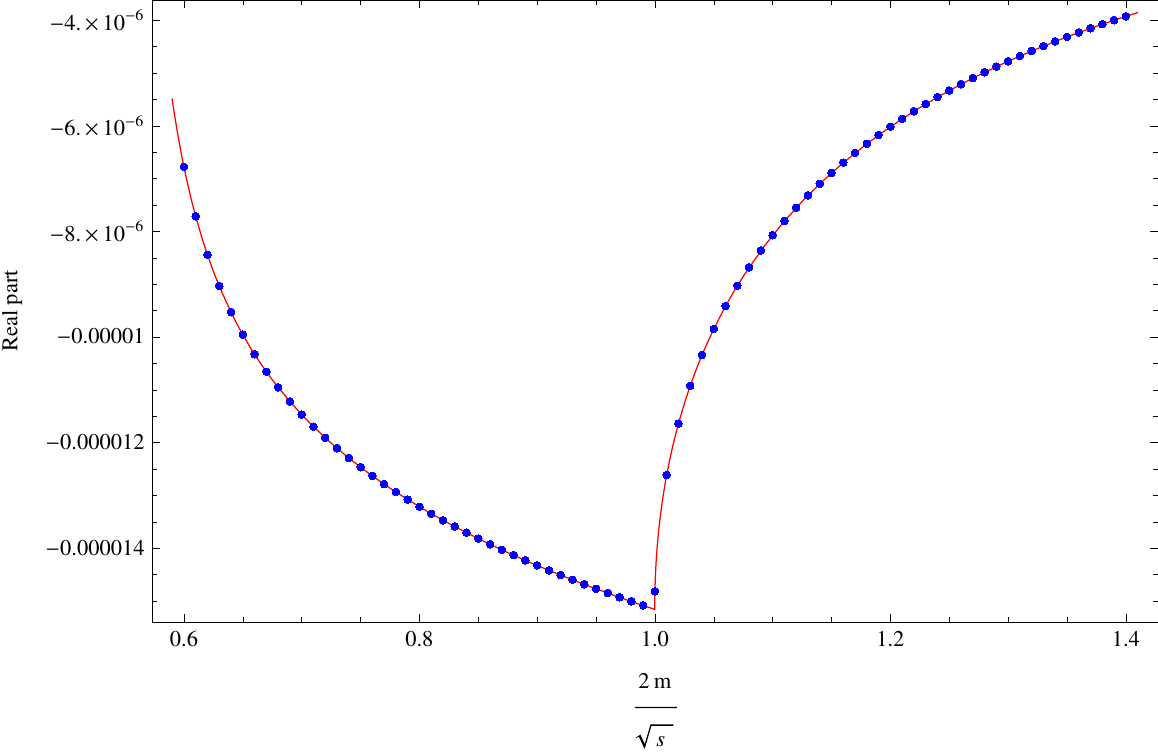}
  \caption{Real part.}
    \label{fig:thre}
  \end{subfigure}
      \begin{subfigure}[h]{0.49\textwidth}
\includegraphics[scale=0.62]{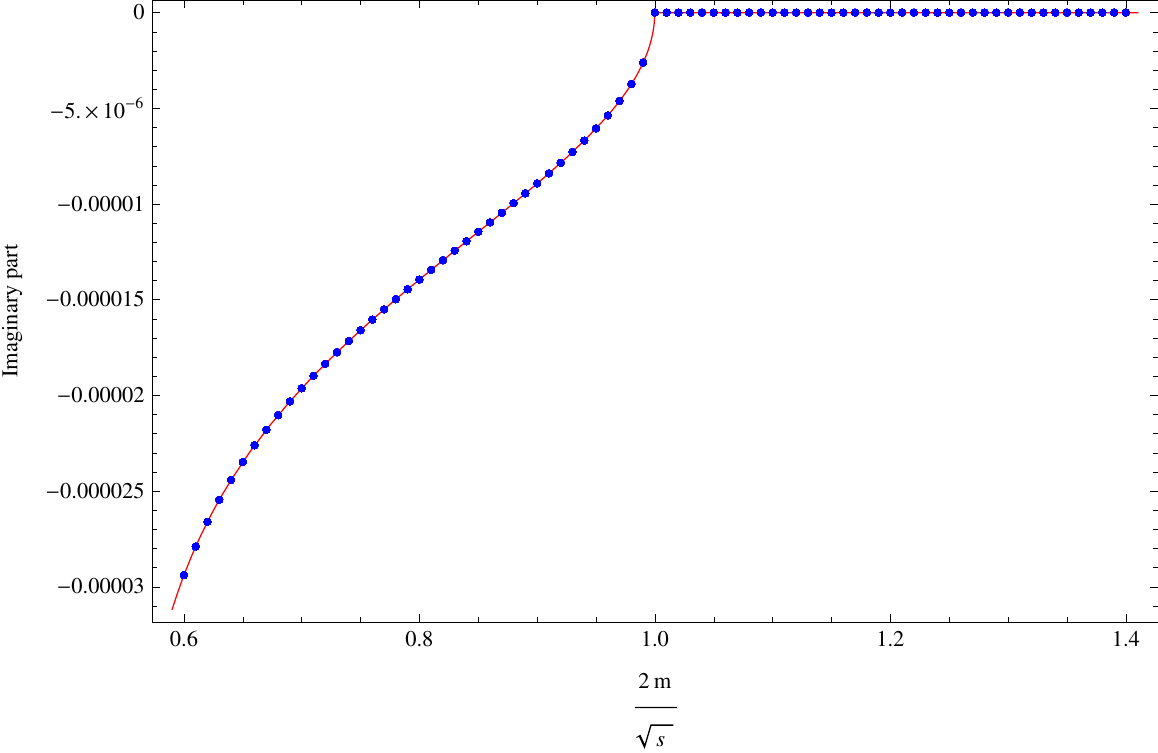}
  \caption{Imaginary part.}
    \label{fig:thrim}
    \end{subfigure}
    \caption{The curve is LoopTools, the dots are the LTD.}
  \end{figure}

\begin{figure}[h]
  \centering
  \begin{subfigure}[h]{0.49\textwidth}
\includegraphics[scale=0.8]{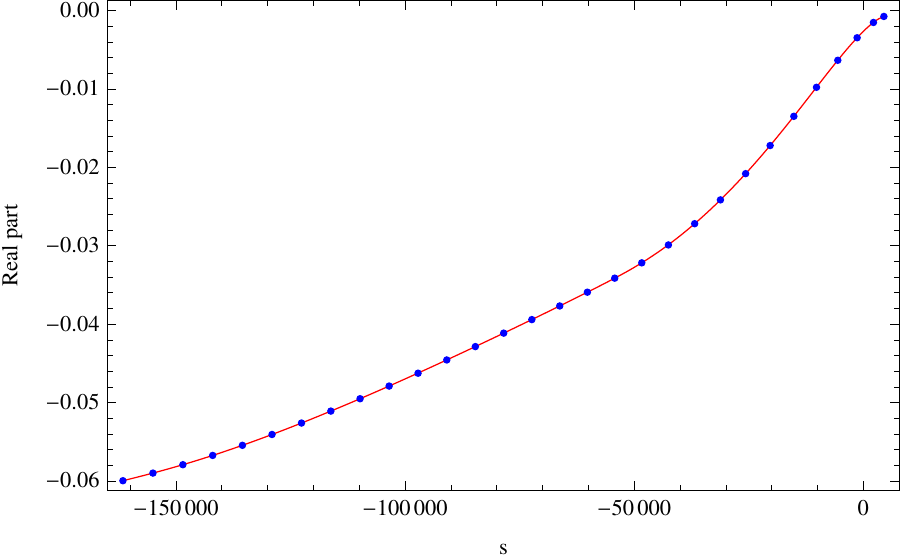}
  \caption{Real part.}
  \end{subfigure}
      \begin{subfigure}[h]{0.49\textwidth}
\includegraphics[scale=0.8]{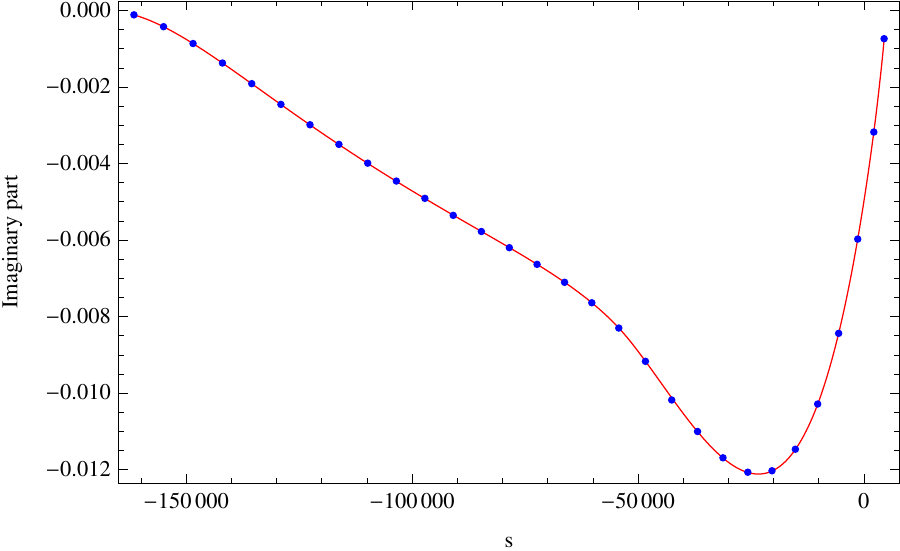}
  \caption{Imaginary part.}
    \end{subfigure}
    \caption{The curve is LoopTools, the dots are the LTD.}
        \label{fig:pentascan}
  \end{figure}

\nin
Our implementation is written in C++ \cite{Buchta:2015xda,Buchta:2015wna}. We use Cuhre from the Cuba-Library \cite{Hahn:2004fe} as a numerical integrator and LoopTools (LT) \cite{Hahn:1999wr} to generate analytic values for comparison. The code runs on an Intel i7 desktop machine with 3.4GHz. It is capable of calculating triangles, boxes and pentagons with no deformation needed with 4 digits precision in 0.5s. The table below shows an explicit result where a pentagon with deformation was calculated with 4 digits precision in 40s.
\begin{table}[H]
\centering
\setlength{\tabcolsep}{0.3pc}
    {\small
\begin{tabular}{|lll|}
\hline
 &Real part & Imaginary part  \\
\hline
LT & $4.643378\times 10^{-14}$  & $-i~1.654437\times 10^{-14}$ \\
LTD & $4.643400(234)\times 10^{-14}$ &  $-i~1.654457(234)\times 10^{-14}$ \\
\hline
\end{tabular}
}
\label{tab:pentascalar}
\end{table}
Another test the program has been put through is a scan of the region around threshold. To produce the two graphs in Fig. \ref{fig:thre} and \ref{fig:thrim}, a triangle has been taken; the center-of-mass energy $s$ has been kept constant while the mass $m$ (which was taken to be identical for all three internal lines) was varied.
From the plots one can see that LoopTools' values are well matched by the computer program. In particular, there is no loss of precision or increase in calculation time close to threshold. 
The LTD relation for scalar loop integrals can easily be extended to deal with 
tensor integrals in a straightforward manner.  
If the one-loop integral features a non-trivial numerator $\mathcal{N}(\ell,\{p_i\})$,   
the LTD theorem takes the form
\bea
L^{(1)}(p_1, \dots, p_N; \mathcal{N}(\ell,\{p_i\})) 
= - \sum_{i\in\alpha_1} \, \int_{\ell} \, \td(q_i) \,
\mathcal{N}(\ell,\{p_i\}) \,
\prod_{\substack{j\in\alpha_1 \\ j\neq i}} \,G_D(q_i;q_j)~.
\eea   
While the numerator is formally left unchanged, the dual delta function 
demands $q_{i,0}^{(+)}=\sqrt{\mathbf{q}_i^2+m_i^2-i0}$, which is equivalent to 
$\ell_{0}=-k_{i,0}+\sqrt{\mathbf{q}_i^2+m_i^2-i0}$.
As a direct consequence, the numerator 
takes a different form in each dual contribution. 
The cancellation of 
singularities among dual contributions is left intact. 
Below, we give an example of a hexagon sample point (reference point P. 32 in \cite{Buchta:2015xda} or P24 in \cite{Buchta:2015wna}) with a rank 3 numerator and all 
internal masses different which is calculated in about 85s.
\begin{table}[H]
\centering
\setlength{\tabcolsep}{0.3pc}
    {\small
\begin{tabular}{|lll|}
\hline
 &Real part & Imaginary part  \\
\hline
SecDec & $-2.07531(19) \times 10^{-6}$ & $+ i~6.97158(56) \times 10^{-7}$ \\
LTD & $-2.07526(8)  \times 10^{-6}$ & $+ i~6.97192(8) \times 10^{-7}$ \\
\hline
\end{tabular}
}
\label{tab:param}
\end{table}
Since LoopTools is limited to a maximum of 5 external legs, we used SecDec \cite{Borowka:2015mxa} to produce reference values. 
We also did scans to check a broader slice of phase space. Fig. \ref{fig:pentascan} shows the performance of the code when varying the momentum $p_1$ and thus the center of mass energy $s$.
For this scan we chose the numerator $(\ell\cdot p_1)\times (\ell\cdot p_2)\times (\ell\cdot p_3)$. Hence, both numerator and 
denominator are varied simultaneously. The program passed this test successfully, as well.\\
Despite still being in development, the program already shows competitive speed and precision when calculating loop integrals. We have demonstrated that it deals well with scalar and also tensor graphs up to the hexagon level. An extension to more external legs is straightforward and easy to realize.
\section*{Acknowledgements}
\nin
This work has been supported by the Research Executive Agency (REA)
of the European Union under the Grant Agreement number PITN-
GA-2010-264564 (LHCPhenoNet), by
the Spanish Government and EU ERDF funds (grants FPA2014-53631-C2-1-P, FPA2011-23778 and CSD2007-00042 Consolider
Project CPAN) and by GV (PROMETEUII/2013/007). SB acknowledges support from JAEPre programme (CSIC).

\end{document}